\documentclass[apj]{emulateapj}
\usepackage{mathptmx}

\def\gtorder{\mathrel{\raise.3ex\hbox{$>$}\mkern-14mu
             \lower0.6ex\hbox{$\sim$}}}
\def\ltorder{\mathrel{\raise.3ex\hbox{$<$}\mkern-14mu
             \lower0.6ex\hbox{$\sim$}}}

\slugcomment{Submitted to ApJ}

\shorttitle{FRB persistent sources}

\shortauthors{Ofek}

\begin{document}




\title{A search for FRB\,121102-like persistent radio-luminous sources -- Candidates and implications for the FRB rate and searches}
\author{Eran O. Ofek\altaffilmark{1}
}
\altaffiltext{1}{Benoziyo Center for Astrophysics, Weizmann Institute
  of Science, 76100 Rehovot, Israel}

\begin{abstract}

The localization of the repeating fast radio burst (FRB), FRB\,121102,
suggests that it is associated with a persistent radio-luminous compact source
in the FRB host galaxy.
Using the FIRST radio catalog, I present a search for 
luminous persistent sources in nearby galaxies,
with radio luminosities $>10\%$ of the FRB\,121102
persistent source luminosity.
The galaxy sample
contains about 30\% of the total galaxy
$g$-band luminosity within $<108$\,Mpc,
in a footprint of 10,600\,deg$^{2}$.
After rejecting sources likely due to active galactic nuclei
activity or background sources,
I remain with 11 candidates that are presumably associated with galactic disks
or star formation regions.
At least some of these candidates are likely to be due to chance alignment.
In addition, I find 85 sources within $1''$ of galactic nuclei.
Assuming the radio persistent sources are not related
to galactic nuclei and that they follow the galaxy $g$-band light,
the 11 sources imply a 95\% confidence upper limit on the space density of
luminous persistent sources of $\ltorder5\times10^{-5}$\,Mpc$^{-3}$,
and that at any given time only a small fraction
of galaxies host a radio luminous persistent source
($\ltorder10^{-3}$\,$L_{*}^{-1}$).
Assuming persistent sources life time of 100\,yr, this implies
a birth rate of $\ltorder5\times10^{-7}$\,yr$^{-1}$\,Mpc$^{-3}$.
Given the FRB volumetric rate, and
assuming that all FRBs repeat and are associated with persistent radio sources,
this sets a lower limit on the rate of FRB events per persistent source
of $\gtorder0.8$\,yr$^{-1}$.
I argue that these 11 candidates are good targets for FRB searches
and I estimate the FRB event rate from these candidates.

\end{abstract}

\keywords{
galaxies: general ---
galaxies: statistics}

\section{Introduction}
\label{sec:Introduction}

Fast radio bursts are dispersed millisecond-duration pulses observed
at GHz frequencies (Lorimer et al. 2007; Thornton et al. 2013).
Spitler et al. (2016) reported the discovery of a repeating FRB,
and follow-up observations using
the Very Large Array (VLA) were able to provide an arcsecond localization
of the source of the bursts (Chatterjee et al. 2017).
The localization is consistent with the position of a
star formation region in a faint galaxy ($r=25.1$\,mag)
at a redshift of $0.19273$ (Tendulkar et al. 2017; Bassa et al. 2017).
Furthermore, Chatterjee et al. (2017) reported the existence of a persistent radio source
within $1''$ of the FRB location,
and Marcote et al. (2017) showed that the
FRB and the persistent source are separated by less than 12\,mas
($\ltorder40$\,pc projected).
The radio source has a mean flux density of about $0.18$\,mJy,
it is variable, presumably due to scintillations,
and unresolved with an angular size smaller than about 1.7\,mas.

There are many suggestions for the FRB mechanism,
including soft gamma-ray repeats
(e.g., Popov \& Postnov 2010;
Kulkarni et al. 2014;
Lyubarsky 2014; see however Tendulkar et al. 2016),
Galactic stars (e.g., Maoz et al. 2015),
pulsars (e.g., Lyutikov et al. 2016; Katz 2017a),
magnetars (e.g., Metzger et al. 2017),
pulsar wind nebulae (e.g., Dai, Wang \& Yu 2017),
active galactic nuclei (Katz 2017b), and more.
Based on the FRB\,121102 persistent source luminosity, spectrum, angular size,
and the FRBs constant dispersion measure (DM) over a year time scale,
Waxman (2017) inferred the
properties of the emitting region.
He concluded it is a mildly relativistic,
$10^{-5}$\,M$_{\odot}$ shell which propagates into an ambient medium.
The energy of this expanding shell is of the order of $10^{49}$\,erg,
and its lifetime is of order $\ltorder300$\,yr.
Waxman (2017) also suggested an efficient mechanism that produces
the FRBs within this persistent source via synchrotron maser.

An intriguing property of the FRB\,121102 persistent source is that it is
radio bright.
With an isotropic radiative energy of $\nu L_{\nu}\approx3\times10^{38}$\,erg\,s$^{-1}$,
at $1.4$\,GHz,
it is over an order of magnitude brighter
than the brightest known supernova (SN) remnants
(e.g., Lonsdale et al. 2006; Parra et al. 2007; Chomiuk \& Wilcots 2009).
However, its luminosity is comparable with the brightest
young ($\sim1$\,yr old) SNe (e.g., Weiler et al. 2002)

Spitler et al. (2016) showed that the FRBs' arrival times are highly
non Poissonic (see also e.g., Wang \& Yu 2017).
This fact makes it difficult to estimate the actual mean rate of the
FRB\,121102 events.

Here, I report on a search for persistent radio sources in galaxies in the nearby Universe ($z<0.025$).
I find 11 candidates, with luminosity greater than 10\% of
the FRB121102 persistent source.
Regardless of the nature of these luminous radio sources,
and their relation to FRBs,
by estimating the completeness of the survey I was able to place an upper limit
on the number density of such bright compact persistent sources
in the local Universe.
Furthermore, assuming that all the FRBs repeat and
associated with persistent radio sources,
I set a lower limit on the rate of FRB events per persistent source.
I also discuss the volumetric rate of FRBs
and the implications for FRB searches.

The structure of this paper is as follows:
In \S\ref{sec:search} I describe the search for persistent sources
in the nearby Universe, while in \S\ref{sec:analysis} I analyze
the results.
The nature of the persistent radio sources is discussed in \S\ref{sec:Nature},
and in \S\ref{sec:disc} I summarize the results and discuss the
implications for FRB searches.

\section{Search for FRB persistent sources}
\label{sec:search}

Here I present a search for persistent radio-luminous compact sources
in the nearby Universe.
In \S\ref{sec:gal} I present the nearby galaxy sample,
and in \S\ref{sec:xcorr} I describe the
cross matching of the galaxy list with the
FIRST radio catalog.
In \S\ref{sec:Chance} I estimate chance coincidence probabilities,
while in \S\ref{sec:Var} I discuss the radio sources' variability.
All the steps and analysis were performed using tools and catalogue available in
the MATLAB Astronomy \& Astrophysics
Toolbox\footnote{https://webhome.weizmann.ac.il/home/eofek/matlab/} (Ofek 2014).

\subsection{The nearby galaxy sample and its completeness}
\label{sec:gal}

I compiled a catalog of nearby galaxies within 108\,Mpc.
The catalog is based on combining the HyperLEDA
galaxies\footnote{http://leda.univ-lyon1.fr/}
(Paturel et al. 2003; Makarov et al. 2014)
with the NASA Extragalactic Database (NED\footnote{https://ned.ipac.caltech.edu/})
redshifts, and the Sloan Digital Sky Survey (SDSS; York et al. 2000) galaxies with known redshifts.
Both catalogs are restricted to the FIRST\footnote{FIRST catalog version 2017 Dec 14.}
radio survey footprint (Becker, White, \& Helfand 1995).
This catalog is far from being complete 
and I estimate its completeness below.
I note that the total area of the FIRST footprint is about 10,600\,deg$^{2}$.

The HyperLEDA catalog lists galaxies brighter than 18 mag, without redshifts.
I compiled a catalog of all redshifts available in
NED\footnote{This catalog, with $3.27\times10^{6}$ redshifts, is available as part of the MATALB Astronomy \& Astrophysics Toolbox (Ofek 2014), under {\tt cats.galaxies.NED\_z}.}.
I cross-matched galaxies in the HyperLEDA catalog with the redshift catalog.
The association radius was set to $15''$.
Next, I selected only galaxies with redshifts in the range 0 to 0.025 that are
found in the FIRST footprints.
Specifically, I demand that the FIRST coverage maps at the galaxy position
have rms below 0.26\,mJy and above zero.
The resulting catalog has 16,152 entries.

I supplemented this catalog with SDSS galaxies in the FIRST
footprint\footnote{The SDSS catalog contains many HyperLEDA galaxies.},
with redshifts above 0 and below 0.025 that are not listed in the
HyperLEDA catalog ($5''$ association).
This catalog contains an additional 12,663 galaxies.
All the galaxies' magnitudes were corrected for Galactic extinction
(Cardelli et al. 1989; Schlegel et al. 1998).

\begin{figure}
\centerline{\includegraphics[width=8cm]{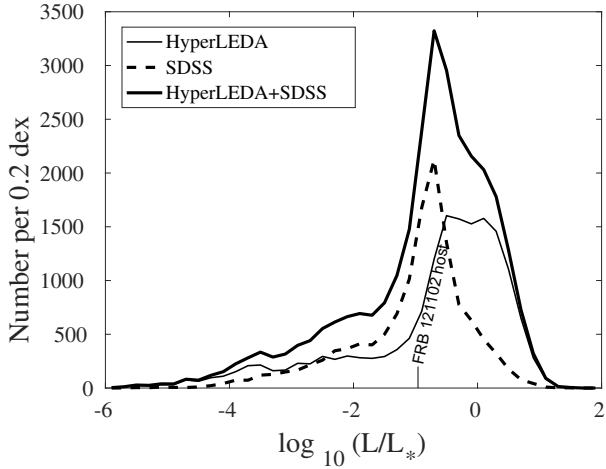}}
\caption{The luminosity function of the HyperLEDA (thin solid line),
SDSS (dashed line), and combined catalog galaxies (thick solid line).
The luminosity is presented in
units of $L_{*}$ which I here define to corresponds to $g$-band
absolute magnitude of $-19$.
All galaxy magnitude are SDSS $g$-band model magnitude.
The Y-axis shows the number of galaxies per 0.2 logarithmic bins.
The vertical line on the bottom indicates the approximate luminosity
of the FRB\,121102 host galaxy.
\label{fig:GalSample_LumFun}}
\end{figure}

The luminosity functions of the two galaxy catalogue
are shown in Figure~\ref{fig:GalSample_LumFun}.
The luminosity function  shows a steep drop for galaxies fainter
than about $0.1$\,$L_{*}$.
I note that the FRB\,121102 host galaxy luminosity is near
the peak of the luminosity function in Figure~\ref{fig:GalSample_LumFun}.

\begin{figure}
\centerline{\includegraphics[width=8cm]{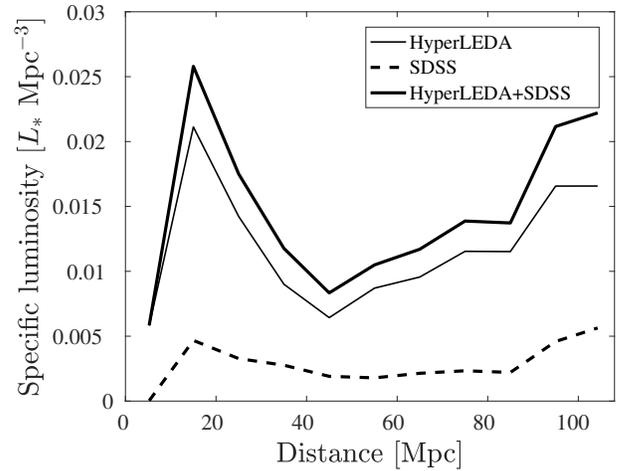}}
\caption{The luminosity density in the galaxy catalog as a function
of distance (corrected for the FIRST footprint sky area).
\label{fig:VolLum_dist}}
\end{figure}

Next, I estimate the completeness of the combined galaxy catalog in terms of total $g$-band
luminosity in the nearby Universe.
Figure~\ref{fig:VolLum_dist} shows the galaxy
sample $g$-band luminosity per unit volume
as a function of distance.
I also integrated the total $B$-band luminosity in the Karachentsev et al. (2004)
catalog of galaxies within 10\,Mpc.
In order to correct for the galaxy zone of avoidance, I restricted the search
again to the FIRST footprints.
Since the Karachentsev et al. catalog lists the $B$-band magnitude, while
our catalog uses the SDSS $g$-band,
I use a $B(Vega)-g(AB)=0.18$\,mag
correction\footnote{Assuming a black-body spectrum with an effective temperature of 6000\,K.}.
I find that the $g$-band luminosity density
in the Karachentsev et al. (2004) catalog
is about 0.055\,$L_{*}$\,Mpc$^{-3}$,
where here $L_{*}$ is the luminosity corresponding to $g$-band absolute magnitude of $-19$.
This is close to the measured $L_{*}$ at $z=0.1$ ($\approx-18.8$;
e.g., Montero-Dorta \& Prada 2009).

Assuming that the Karachentsev et al. (2004) catalog is nearly complete in terms
of the total luminosity and star formation
rate\footnote{If this assumption was wildly incorrect, then we would expect that
blind SN surveys will find many more nearby SN associated with faint dwarf galaxies.}, and that
the galaxy luminosity density within 10\,Mpc is representative
of the galaxy luminosity density at larger distances,
I calculate how much luminosity is missing from of
our HyperLEDA$+$SDSS catalog per unit volume as a function of distance, and integrate.
I find that the total luminosity in our catalog has a completeness of $30\%$.
I note that this factor is uncertain due to the fact that the galaxy distribution is correlated.
Given the galaxy correlation function
(Tucker et al. 1997; Zehavi et al. 2002) I estimate the completeness to
have about 10\% uncertainty.

\subsection{Search for radio sources associated with nearby galaxies}
\label{sec:xcorr}

I cross-matched our galaxy catalogs with FIRST point sources.
I defined point sources to have a major axis smaller than
3 times the uncertainty\footnote{The size uncertainty is calculated using
the formula in: http://sundog.stsci.edu/first/catalogs/readme.html}
in the radio size measurement.
The HyperLEDA-catalog search radius,
for each galaxy, was set to the galaxy's
25\,mag\,arcsec$^{-2}$ surface brightness
semi-major axis listed in the catalog,
while for the SDSS sample I used the SDSS Petrosian
radius\footnote{http://skyserver.sdss.org/dr7/en/help/docs/algorithm.asp?key=mag\_petro}.

In the cross-matching step, I selected only
sources with intrinsic radio luminosity (assuming at the galaxy redshift) of $>10\%$ 
of the luminosity of the persistent source associated with FRB\,121102.
For this, I used the FRB\,121102 persistent source
mean radio flux density of 0.18\,mJy and luminosity distance of 938\,Mpc
(i.e., $\nu L_{\nu}\approx3\times10^{38}$\,erg\,s$^{-1}$).

The cross-matching yielded 122 possible matches, of which
91 are from the HyperLEDA catalog and
31 are from the SDSS catalog.
The candidates are listed in Table~\ref{tab:Cand},
and in \S\ref{sec:Chance} I discuss the chance coincidence probability
for these sources.
\begin{deluxetable*}{llllllllllll}
\tablecolumns{12}
\tablewidth{0pt}
\tablecaption{Luminous persistent radio source candidates}
\tablehead{
\colhead{R.A.}                &
\colhead{Dec.}                &
\colhead{$f_{\rm p}$}          &
\colhead{$\Delta{f_{\rm p}}$}   &
\colhead{$L/L_{\rm pers}$}      &
\colhead{z}                   &
\colhead{$\theta$}            &
\colhead{$f_{\rm NVSS}$}        &
\colhead{$\Delta{f_{\rm NVSS}}$}&
\colhead{$\chi_{{\rm NVSS-FIRST}}$}&
\colhead{$X_{{\rm c}}$}         &
\colhead{Comment}             \\
\colhead{}                    &
\colhead{}                    &
\colhead{mJy}                 &
\colhead{mJy}                 &
\colhead{}                    &
\colhead{}                    &
\colhead{arcsec}              &
\colhead{mJy}                 &
\colhead{mJy}                 &
\colhead{}                    &
\colhead{ct/ks}               &
\colhead{}                    
}
\startdata
09:27:58.282 & $-$02:25:58.95 &     2.1 & 0.14 & 0.14 & 0.023 &  14.32 &    \nodata &  \nodata &   \nodata &   \nodata & Spiral arm + IR source \\ 
10:47:26.693 & $+$06:02:47.72 &     2.9 & 0.14 & 0.13 & 0.019 &   5.87 &    2.4 &  0.4 &  -1.1 &   \nodata             & Off galaxy center; passive galaxy \\ 
23:53:51.412 & $+$07:58:35.91 &     4.2 & 0.13 & 0.16 & 0.018 &  42.68 &    \nodata &  \nodata &   \nodata &   \nodata & Near spiral arm; near red+IR source \\ 
14:10:43.667 & $+$08:59:29.96 &     3.2 & 0.15 & 0.21 & 0.023 &  17.66 &    4.0 &  0.4 &   1.8 &   \nodata             & Edge of spiral disk; red faint source? \\ 
10:25:26.189 & $+$17:15:47.97 &     2.8 & 0.13 & 0.11 & 0.018 &   7.08 &    \nodata &  \nodata &   \nodata &   \nodata & Spiral arm \\ 
10:58:23.641 & $+$24:13:55.32 &     2.3 & 0.15 & 0.12 & 0.021 &  29.79 &    \nodata &  \nodata &   \nodata &   \nodata & Spiral arm \\ 
13:14:41.932 & $+$29:59:59.19 &     2.2 & 0.14 & 0.14 & 0.023 &  20.58 &    4.0 &  0.5 &   3.4 &   \nodata             & Edge of spiral galaxy; IR source \\ 
16:22:44.571 & $+$32:12:59.28 &     2.0 & 0.15 & 0.11 & 0.022 &   0.89 &    2.7 &  0.4 &   1.7 &   \nodata             & Small blue galaxy; near center \\ 
14:00:38.929 & $-$02:51:22.79 &     1.5 & 0.15 & 0.11 & 0.025 &  26.41 &    \nodata &  \nodata &   \nodata &  18.8     & Elliptical galaxy halo; no vis/IR source \\ 
11:45:29.346 & $+$19:23:27.46 &     3.5 & 0.20 & 0.26 & 0.025 &  33.35 &    2.4 &  0.4 &  -2.3 &   \nodata             & Edge of galaxy; No optical or IR source \\ 
14:19:18.855 & $+$39:40:36.03 &    21.1 & 0.15 & 0.95 & 0.020 &   0.50 &   18.5 &  1.0 &  -2.2 &   \nodata             & Compact blue star forming galaxy \\ 
\hline
00:17:27.498 & $-$09:34:26.56 &     2.5 & 0.15 & 0.15 & 0.023 &   0.51 &    \nodata &  \nodata &   \nodata &   \nodata & Center of galaxy \\ 
00:17:59.547 & $-$09:16:00.89 &     1.6 & 0.15 & 0.10 & 0.023 &  19.06 &    2.9 &  0.5 &   2.5 &   \nodata             & Halo of galaxy, likely IR source \\ 
02:52:42.189 & $-$08:48:15.76 &     3.1 & 0.15 & 0.12 & 0.018 &   0.18 &    4.4 &  0.5 &   2.4 &   \nodata             & Center of galaxy \\ 
11:24:03.341 & $-$07:47:01.13 &     2.1 & 0.14 & 0.16 & 0.025 &   0.50 &    3.7 &  0.6 &   2.5 &   \nodata             & Center of galaxy \\ 
01:44:43.099 & $-$04:07:46.25 &     5.2 & 0.13 & 0.21 & 0.018 &  34.81 &   10.9 &  0.6 &   9.0 &   \nodata             & Halo of galaxy, blue source + IR source 
\enddata
\tablecomments{A list of radio sources that spatially coincide with
nearby galaxies.
Sources above the horizontal-line separator are the 11 candidates.
The full table is available electronically, and
here I present only the first several entries.
R.A. and Dec. are the J2000.0 right ascension and declination, respectively,
of the radio source, $f_{\rm p}$ is its peak radio flux, $L/L_{\rm pers}$ is its luminosity
in units of the mean luminosity of the FRB\,121102
persistence source,
$z$ is the spatially coincident galaxy redshift,
and $\theta$ is the angular separation between the galaxy and radio source.
$f_{\rm NVSS}$ and $\Delta{f_{\rm NVSS}}$ are the NVSS flux density and its uncertainty, respectively,
$\chi_{{\rm NVSS-FIRST}}$ is the FIRST to NVSS variability in units of the 1-$\sigma$ uncertainty (Equation~\ref{eq:chi}),
and $X_{{\rm c}}$ is the {\it ROSAT} counts per kilo-second,
from the {\it ROSAT} bright and faint source catalogs (Voges et al. 1999; 2000).
For the ROSAT catalogue, I used a $45''$ search radius.
Entries with no data indicate non detection.
}
\label{tab:Cand}
\end{deluxetable*}
I inspected the SDSS and WISE (Wright et al. 2010) 4.6\,micron-band images of all candidates.
The most common candidates are associated with galactic nuclei to within
1\,arcsec (85 sources). In these cases I assumed that the radio source is due to
active galactic nuclei (AGN) activity in the galaxy center.
As most radio sources above a flux density of a few mJy are
AGNs, this assumption is probably reasonable.
I further note that at a redshift of 0.025
one arcsecond corresponds to
about $0.5$\,kpc, which is an order of magnitude smaller than the
typical size of galaxies. This gives some confidence
that even if we remove from our sample persistent sources
which are spatially coincidence with their galaxy's center,
the number of non-AGN sources missed is small.

\begin{figure*}
\centerline{\includegraphics[width=17cm]{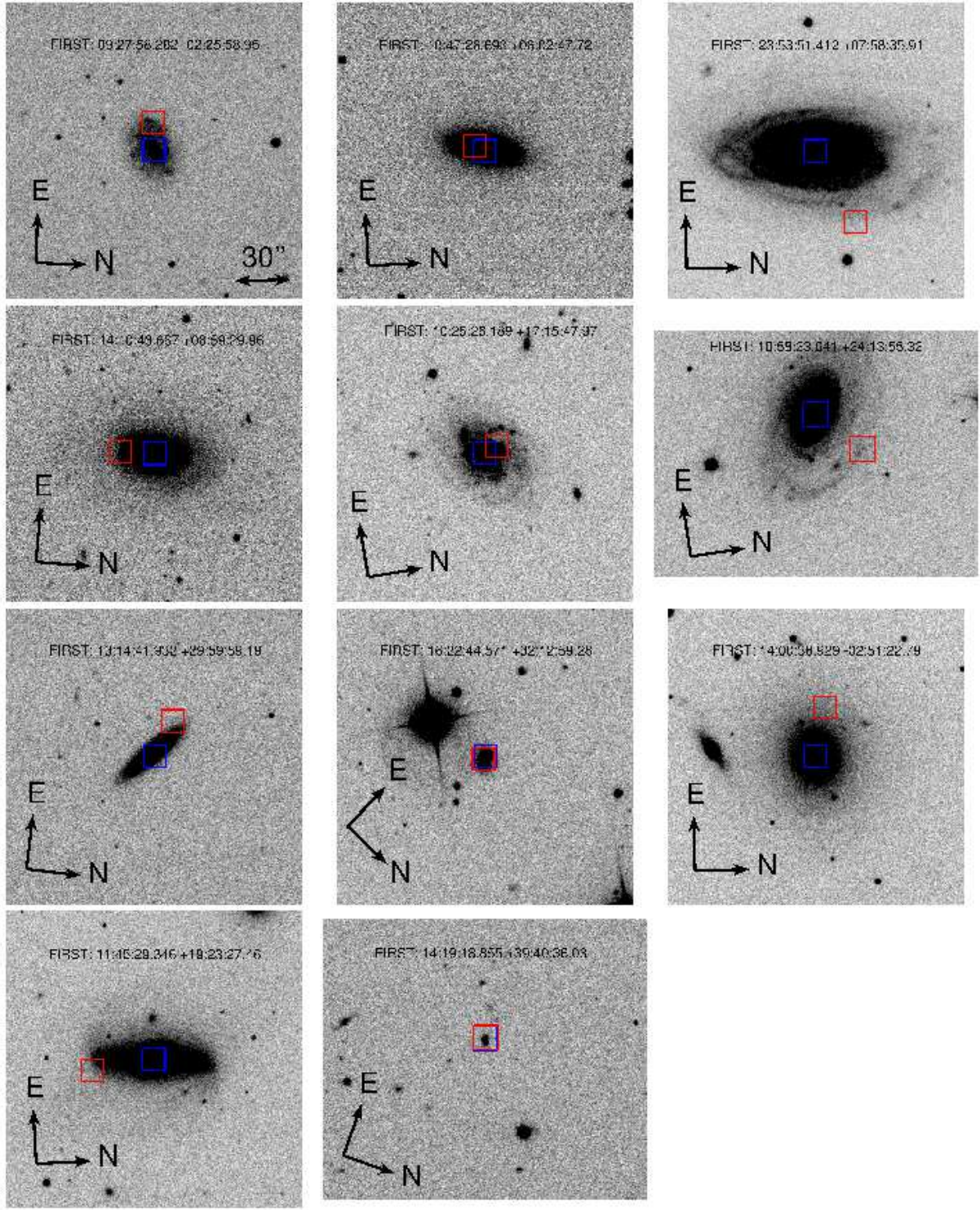}}
\caption{SDSS $g$-band images of the 11 luminous persistent source candidates.
The blue boxes indicate the galaxies center while the red boxes show
the radio sources position. See additional details in Table~\ref{tab:Cand}.
\label{fig:FRB_Pers_Cand}}
\end{figure*}

Many other sources seem to be projected on a galaxy, but outside
any star formation region or light associated with the galaxy -- these sources
are sometimes associated with an unresolved source (likely a background quasar).
Finally, 11 sources were found to coincide with galactic disks light,
or compact star forming galaxies
(similar to FRB\,121102; Bassa et al. 2017).
I regard these 11 sources as the persistent-source candidates,
and they are listed at the top part of Table~\ref{tab:Cand}.
However, if FRBs are related to AGN activity (e.g., Katz 2017; Vieyro et al. 2017), then the number of
candidates changes to 85.
Figure~\ref{fig:FRB_Pers_Cand} shows the SDSS images of the 11 candidate
galaxies,
with markers showing the radio source position and galaxy center.

\subsection{Chance coincidence}
\label{sec:Chance}

Given the total sky area of galaxies in our catalogs ($\cong5.3$\,deg$^{2}$),
it is likely that at least some of the candidates are due to
chance coincidence of background high-redshift sources with
nearby galaxies. 
Here, I estimate the probability that the sources I find
are background sources, unrelated to the spatially associated
low-redshift galaxy.

For each galaxy in the catalogs (HyperLEDA and SDSS) I calculate
its area on the celestial sphere (given its radius).
Furthermore, given the galaxy distance, I calculate
the number density of sources on the sky that are compact
radio sources with luminosity $>10\%$ of the FRB\,121102
persistent radio-source luminosity.
By multiplying the area of each galaxy by the corresponding surface density of
radio sources and summing, I find that the expectancy value for the number
of chance-coincidence background sources is 37.7.
Excluding the 85 sources found within 1'' from galactic nuclei,
there are 37 sources spatially associated with nearby galaxies
(i.e., $=122-85$). The $1''$ exclusion is small
compared with the typical galaxy dimension and has a negligible
effect on the expectancy value for chance coincidence.

Given that the expectancy is 37.7 and that the observed number is 37,
it is likely that at least some of the 11 candidates are due to
chance coincidence.
Assuming that the selection process (e.g., association with galaxy light in images)
selects the best candidates, than
the probability that all the 11 candidates are
associated with their galaxy (rather than background sources)
is 2.9\%.
The probability that at most 5, 2, 1, 0 of the 11 candidates are
associated with their galaxy is 20\%, 40\%, 43\%, and 50\%,
respectively.

This analysis suggests that some, or even all, of the 11 candidates
I found are background sources unrelated to the spatially associated
galaxy.
If follow-up observations will indicate that all 11 candidates
are unrelated to FRBs this will improve the upper limit on the
luminous radio source space density (see \S\ref{sec:perden})
by a factor of about 7.

I note that a reasonable follow-up prioritization of the 11 candidates
in Table~\ref{tab:Cand} is by the inverse galaxy size
and association with star-forming regions.
One reason is that smaller galaxies have lower probability
for chance coincidence with background objects.
In this respect, the most interesting candidate in the list
is J141918$+$394036.
This source has the highest luminosity of all candidates and it is associated
with a small-area blue galaxy.

\subsection{Source variability}
\label{sec:Var}

I cross-matched the sources in Table~\ref{tab:Cand} with the NVSS catalog (Condon et al. 1998)
with a $15''$ match radius.
The NVSS flux and error are listed in Table~\ref{tab:Cand}.
I further calculated, and list in Table~\ref{tab:Cand},
\begin{equation}
\chi_{\rm NVSS - FIRST} = \frac{f_{{\rm NVSS}} - f_{{\rm FIRST}}}{\sqrt{\Delta{f_{{\rm NVSS}}}^2 + \Delta{f_{{\rm FIRST}}}^2 + (\epsilon_{{\rm cos}}f_{{\rm FIRST}})^2 }}.
\label{eq:chi}
\end{equation}
Here $f_{{\rm NVSS}}$ is the NVSS\footnote{The NVSS and FIRST measurements each give a weighted mean flux over several epochs.}
peak-flux density, $f_{{\rm FIRST}}$ is the FIRST peak-flux density,
$\Delta{f_{{\rm NVSS}}}$, and $\Delta{f_{{\rm FIRST}}}$ are the NVSS and FIRST flux errors, respectively,
and $\epsilon_{{\rm cos}}$ is a calibration error assumed to be $0.03$ (Condon et al. 1998).
I note that Ofek et al. (2011) measurements of a few calibration sources suggest that the 
VLA calibration error may be smaller.
Furthermore, Ofek \& Frail (2011) found that any
systematic offset between the FIRST and NVSS fluxes
is small compared with the typical flux errors.

The FIRST-NVSS variability search indicates that these sources are roughly constant.
However, I cannot rule out small amplitude variability due to scintillation
or some long term decrease or increase in flux.

\section{Analysis}
\label{sec:analysis}

\subsection{Persistent source volumetric density}
\label{sec:perden}

\begin{figure}
\centerline{\includegraphics[width=8cm]{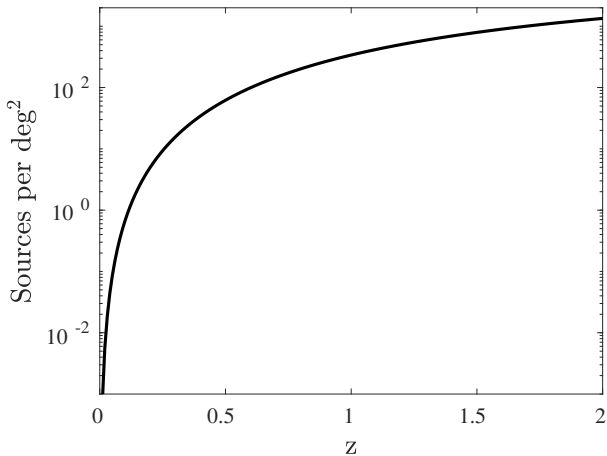}}
\caption{The upper limit on the sky
surface density of radio luminous persistent sources up to a given redshift.
\label{fig:PersRadioSource_SurfaceDensity}}
\end{figure}

I found 11 persistent source candidates. This sets a 95\% (1-sided)
upper limit of 18.2 sources in the searched volume (Gehrels 1986).
Assuming the persistent source density is proportional to the
$g$-band luminosity, and 
given the sky area and completeness of the catalog (\S\ref{sec:gal}),
this gives 95\% confidence-level upper limit on the number density of luminous persistent sources
of:
\begin{equation}
\rho_{pers}\ltorder5\times10^{-5}\,{\rm Mpc}^{-3}.
\label{eq:Rvp}
\end{equation}
Using the total galaxy luminosity per unit volume I found in \S\ref{sec:gal},
this is equivalent to a number of persistent sources
per $L_{*}$ galaxy of $\ltorder10^{-3}$\,$L_{*}^{-1}$.
However, if FRBs are related to galactic nuclei,
then I find 85 candidates and the limit in Equation~\ref{eq:Rvp}
will change to $\ltorder3\times10^{-4}$\,Mpc$^{-3}$
($\ltorder5\times10^{-3}$\,L$_{*}^{-1}$).
I note that this estimate ignores any possible evolution of persistent sources with redshift.
However, the expected increase in the star formation rate between $z=0$ to $z=0.025$ is only 8\%
(e.g., Wyder et al. 2005; Schiminovich et al. 2005).
Given the density in Equation~\ref{eq:Rvp},
Figure~\ref{fig:PersRadioSource_SurfaceDensity}
presents the upper limit on the sky
surface density of radio luminous persistent sources, as a function
of redshift.

It is not yet clear whether the assumption that FRBs follow the galaxy light
is correct.
For example, based on the low luminosity of the FRB\,121102 host (extinction corrected $r$-band abs. mag. $-16.8$),
Nicholl et al. (2017) claimed that FRBs prefer dwarf galaxies.
Using the Karachentsev et al. (2004) catalog, I calculate the fraction of $B$-band luminosity
in galaxies fainter than the host galaxy of FRB\,121102 (i.e., $g$-band abs. mag. $-16.6$; assuming $g-r\approx0.2$\,mag).
I find that $\approx 3\%$ of the luminosity is in galaxies fainter than abs. mag. $-16.6$.
Since it is expected that there is some incompleteness even in the Karachentsev et al. catalog,
the actual fraction can be a little bit higher.
Therefore, I cannot rule out the possibility
that the FRB\,121102 host galaxy luminosity follows the general
galaxy luminosity function, at more than 97\% confidence.

I note that the luminosity of the FRB\,121102 host galaxy
is near the peak of the luminosity function of galaxies
in my catalog.
Therefore, even if the assumption that
persistent radio sources follow the $g$-band light
is incorrect, then my upper limit on the space density
of radio persistent sources is likely still valid
to an order of magnitude.

The upper limit on the FRB\,121102 persistent source age estimated
by Waxman (2017; see also Nicholl et al. 2017) is 300\,yr,
while the lower limit on its age is $\gtorder5$\,yr
(Metzger et al. 2017; Waxman 2017).
Adopting the persistent source age of $\tau_{\rm pers}\approx100$\,yr,
our number density implies a birth rate of 
\begin{equation}
\dot{\rho}_{pers,birthrate} \ltorder 5\times10^{-7} \Big(\frac{t_{\rm age}}{100\,{\rm yr}}\Big)^{-1}\,{\rm yr}^{-1}\,{\rm Mpc}^{-3}.
\end{equation}
This suggests that the origin of FRB persistent sources is some sort of rare
phenomenon (e.g., explosions).
It is tempting to relate this birth rate to that of known events
(e.g., Super Luminous Supernovae [SLSN] or Gamma-Ray Bursts [GRB];
e.g., Nicholl et al. 2017).
Howevre, SLSN and GRB can be seen to large distances, and it is very likely
that other fainter, yet unknown, transient classes exist.
In this context I note that Waxman (2017)
found that the total energy of the nebula associated with the FRB\,121102 persistent source
is of the order of $10^{49}$\,erg.
This is low relative to the energetics of {\it known} rare events.

\subsection{The rate of FRBs}
\label{sec:frbrate}

In order to constrain the rate of FRB events per persistent source,
we need an estimate of the FRBs rate.
There are many FRB rate estimates (e.g., Deneva et al. 2009; Burke-Spolaor et al. 2014;
Champion et al. 2016; Lawrence et al. 2016; Vander Wiel et al. 2016).
They are typically reported for different parameters (e.g., fluence limit and FRB duration),
and therefore a comparison between these rates requires caution.

Furthermore, I note that FRB searches may be slightly biased by several reasons. For example:
(1) Usually FRB searches are done up to some limiting dispersion measure (DM).
The DM threshold may evolve with time (e.g., whenever a new record in
DM is found, the DM threshold is updated). This may result in an incompleteness
for bright/far FRB events.
(2) FRB searches are performed using non-coherent de-dispersion.
This may bias against FRBs with durations shorter than a fraction of a millisecond.
I note that in Zackay \& Ofek (2016) we suggested a computationally efficient method
for calculating the coherent de-dispersion.

In order to avoid some of these complications,
I prefer an estimate based on a
single instrument.
Therefore, for the rate estimate I adopt the Cahmpion et al. (2016)
analysis of ten FRBs detected by the Parkes telescope.
Cahmpion et al. (2016) report an FRB all-sky rate of
$R(F>0.13\,{\rm mJy\,ms})=7000_{-3000}^{+5000}$\,day$^{-1}$ (95\% confidence errors)
above fluence of 0.13\,Jy\,ms for a minimal 0.128\,ms pulse duration.

I note that since the radio telescope beam is not uniform and the observed cumulative flux density function
of sources (so called $\log(N)$--$\log(S)$) is some power-law
($-3/2$ for Euclidean Universe; see however Vedantham et al. 2016),
this may affect the effective beam size of the survey (used in the rate calculation).
Assuming a cumulative flux density power-law of $-3/2$ and assuming a Gaussian beam with
a size equal to the beam full-width half maximum, the correction factor is 1 (see appendix C in Ofek et al. 2011).
Therefore, here I adopt the Cahmpion et al. (2016) rate.

\subsection{The FRB volumetric rate}
\label{sec:FRBvr}

Assuming that all FRBs repeat, that they
are associated with radio persistent sources,
their emission is isotropic, and using
the FRB rate (\S\ref{sec:frbrate}),
I use the upper limit on the persistent source number density
(Eq.~\ref{eq:Rvp})
to put a lower limit on the rate of FRB events per persistent source.

\begin{figure}
\centerline{\includegraphics[width=8cm]{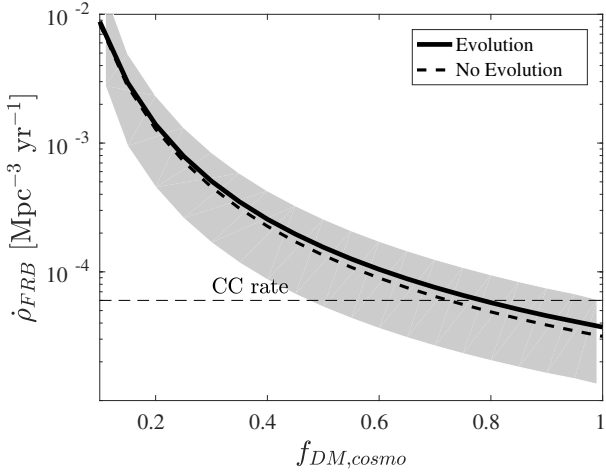}}
\caption{The estimated FRB rate per Mpc$^{-3}$ as a function of $f_{\rm DM,cosmo}$.
The solid line shows the approximate FRB rate density assuming they follow the star formation
rate, while the dashed line is assuming no evolution.
The gray region indicates the approximate 68\% confidence region,
while the dashed horizontal line shows the core collapse SN rate
(e.g., Li et al. 2011).
\label{fig:VolR_FRB_vs_f_cosmo_DM}}
\end{figure}

The first step is to estimate the volumetric rate of FRBs.
To estimate the effective volume of the Parkes search,
I use all the 16 FRBs found by the Parkes radio
telescope\footnote{Adopted from the FRB catalog: http://www.astronomy.swin.edu.au/pulsar/frbcat/}.
For each FRB DM I removed the estimated Milky Way DM,
and I attributed a fraction\footnote{Attributing a fraction of the DM to intergalactic dispersion
is an approximation -- in practice the host galaxy DM is coming from some unknown probability distribution
of host-galaxy DM.}
$f_{\rm DM,cosmo}$ of the remaining DM
to intergalactic dispersion.
The analysis is performed
for $f_{\rm DM,cosmo}$ in the range of $0.1$ to $1$.
I converted the remaining intergalactic DM to redshift via the formulae
in Zheng et al. (2014) assuming Planck 
cosmological parameters (Ade et al. 2015).
Each redshift was transformed to a cosmological volume.
Given a flux limited survey,
Euclidian universe,
and no cosmological evolution, the expectation value of the survey volume
is two times the average of the volumes enclosed by the sources
(e.g., similar to the argument of the $V/V_{max}$ test; e.g., Schmidt 1968).
Given the Universe geometry and
cosmological evolution in the
star formation rate, the ratio between the effective survey volume
and the mean volume (corresponding to the redshift) of sources is somewhat smaller than 2
(i.e., star formation increases with redshift).
In the calculation I assume the FRB rate follows the star formation rate and
take into account the star-formation evolution 
from the compilation based on\footnote{https://ned.ipac.caltech.edu/level5/March14/Madau/Madau5.html}
Wyder et al. (2005), Schiminovich et al. (2005), Robotham \& Driver (2011), and Cucciati et al. (2012).
In any case, the effect of evolution is considerably smaller than the uncertainty
in the rate estimate.
This allows us to calculate the 
FRB volumetric rate as a function of $f_{\rm DM,cosmo}$.
Figure~\ref{fig:VolR_FRB_vs_f_cosmo_DM}
shows the FRB rate as a function of $f_{\rm DM,cosmo}$.
Regardless of $f_{\rm DM,cosmo}$, and assuming
cosmological evolution in the FRB rate that follows the star formation rate,
I find an FRB volumetric rate of
\begin{equation}
\dot{\rho}_{FRB}\gtorder (3.7\pm2.4) \times10^{-5}\,{\rm Mpc}^{-3}\,{\rm yr}^{-1}.
\label{eq:NVfrb}
\end{equation}
This estimate takes into account the effect of cosmological time dilation
by multiplying the observed rate by $(1+<z>)$, where $<z>$ is the
volume-weighted mean redshift of the FRBs.
I note that this estimate depends on the unknown luminosity function of FRBs
and should be regarded as an order of magnitude estimate.
Furthermore, if the FRB emission is beamed then the rate at Equation~\ref{eq:NVfrb}
is, again, only a lower limit.

\subsection{FRB rate per persistent source}
\label{sec:RatePerP}

\begin{figure}
\centerline{\includegraphics[width=8cm]{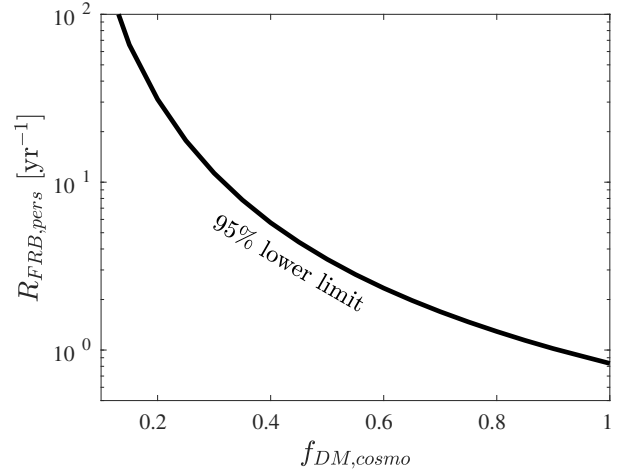}}
\caption{Lower limit on the FRB rate per persistent source as a function of $f_{\rm DM,cosmo}$,
as derived from the upper limit on the FRB persistent sources number density
and the FRB volumetric rate.
This lower limit on the rate corresponds to FRB events
for which the intrinsic luminosity is as high as the FRBs seen by the Parkes telescope.
\label{fig:R_FRB_persist_vs_f_cosmo_DM}}
\end{figure}

Assuming there is a steady state of FRB persistent source density 
that follows the star foramtion rate,
and assuming that all FRBs are associated with
persistent sources,
I divide the lower limit on the FRB rate per unit volume
(Eq.~\ref{eq:NVfrb}) by the upper limit on the persistent sources
space density to derive a lower limit on the rate of FRBs per persistent source.
Figure~\ref{fig:R_FRB_persist_vs_f_cosmo_DM} presents the 95\% confidence
lower limit on rate of FRB events per persistent source
as a function of $f_{\rm DM,cosmo}$.
For $f_{\rm DM,cosmo}=1$ 
I find an FRB rate per persistent source of $R_{FRB,pers}\gtorder0.8$\,yr$^{-1}$.
This lower limit on the rate corresponds to
FRB events for which the intrinsic luminosity
is as high as the FRBs seen by Parkes.


\subsection{Predictions and implications for observing strategy}
\label{sec:obs}

\begin{figure}
\centerline{\includegraphics[width=8cm]{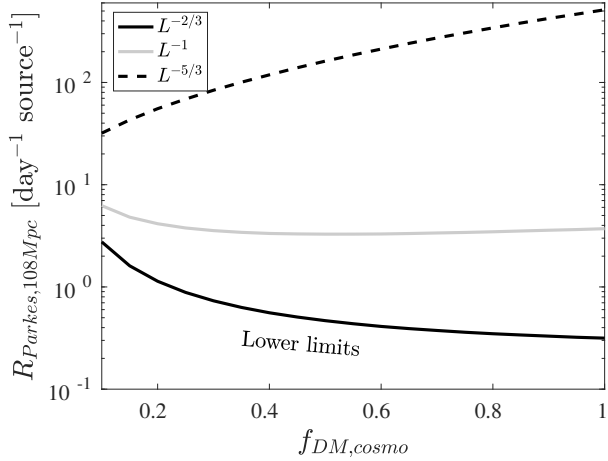}}
\caption{The predicted mean number of FRB events per day per persistent source
detectable
using a Parkes-like telescope, if directed to an FRB-emitter source
(i.e., presumably a persistent radio source) at distance of 108\,Mpc.
This plot is shown for FRB cumulative luminosity function 
$\propto L^{-2/3}$ (solid line), $L^{-1}$ (gray line), and $L^{-5/3}$ (dashed line).
I note that the observed cummulative luminosity function of FRB\,121102 is consistent
with $\propto L^{-2/3}$ (e.g., Nicholl et al. 2017).
\label{fig:PredParks_EventRate_LumFun}}
\end{figure}

An interesting implication is that if all FRBs repeat and are associated
with luminous radio sources, then searches for FRBs in nearby galaxies
(e.g., M31) using small radio dishes
are likely to fail as on average only one in $\gtorder1000$ galaxies
hosts a luminous radio source.
An observing strategy that is favored by my findings
is to monitor for FRBs among the 11 candidates listed in Table~\ref{tab:Cand}.
Figure~\ref{fig:PredParks_EventRate_LumFun} shows the predicted mean number of FRB events
per persistent source per day that may be detected
using a Parkes-like telescope if directed to an FRB-emitter source
(i.e., presumably a persistent radio source) at a distance of 108\,Mpc.
This plot is shown for an FRB cumulative
luminosity function $\propto L^{-2/3}$, $L^{-1}$, and $L^{-5/3}$.
The FRB rate per such source is estimated by
\begin{equation}
R_{\rm Parkes,108\,Mpc}\gtorder R_{\rm FRB, pers} \Big(\frac{<d_{\rm Parkes}>}{108\,Mpc}\Big)^{-2\gamma}.
\label{eq:Nparks}
\end{equation}
Here $R_{\rm FRB, pers}$ is the lower limit on the FRB rate per
persistent source (\S\ref{sec:RatePerP};
Fig.~\ref{fig:R_FRB_persist_vs_f_cosmo_DM}),
$<d_{\rm Parkes}>$ is the mean distance of the Parkes detected FRBs
(\S\ref{sec:FRBvr}; i.e., $z\approx0.7$ for $f_{\rm DM, cosmo}=1$),
and $\gamma$ is the assumed power-law index of the
FRB cumulative luminosity function. 
%
%
I note that if FRBs emission is beamed in a constant direction
then Figure~\ref{fig:PredParks_EventRate_LumFun} is correct
on average for a population.
However, in this case, some persistent radio sources will show no FRB emission.
Furthermore, it is important to note that since the FRB\,121102 events are not generated
by a Poisson process, large deviations from the average expected
rate are possible.

\section{The nature of the luminous radio sources in nearby galaxies}
\label{sec:Nature}

An intriguing question is what is the origin of the luminous radio sources
reported here?
It is likely that at least some of of these sources are background
objects unrelated to a nearby galaxy.
Better estimate of the fraction of background objects
requires follow-up observations.
However, assuming that at least some of these sources are related
to their spatially coincident galaxy, there are still several physical
explanations.

First, it is possible that these sources are similar in nature to the
radio persistent source associated with FRB\,121102.
This possibility can be further tested using follow-up radio spectral,
temporal and interferometric observations.

Second, I would like to consider the possibility that some of these sources
are just the bright end of the supernova (SN) remnants (SNR)
luminosity function,
or young SN with considerable circumstellar material that convert
their kinetic energy to radiation on short time scales ($\sim1$\,yr).
This hypothesis can be tested by long-term monitoring of the radio sources,
and a search for flux variability.
I note that the brightest known radio SNe (e.g., Weiler et al. 2002)
are as bright as the FRB\,121102 persistent radio source.
However, these SNe are variable on time scales of about one year,
which is not consistent with the non-detection of variability between
the NVSS (when available)
and FIRST epochs which are typically several years apart.

Another possibility is gamma ray bursts (GRBs).
In GRBs the ejecta velocity is much higher and therefore
can produce luminous events (e.g., Levinson et al. 2002).
In fact, in the past, the non-detection of transient radio sources
was used to set an upper limit on the GRB space density, which can be translated to
a lower limit on the GRBs beaming factor (Levinson et al. 2002; Gal-Yam et al. 2006).

Levinson et al. (2002) estimated the number of GRB radio afterglows
in a flux-limited survey.
However, our survey is both volume limited and luminosity limited.
Therefore an upper limit on the number of expected GRB afterglows
in our survey is given by their rate in a volume limited survey
\begin{equation}
R_{GRB} \ltorder 0.2 \Big(\frac{d}{108\,{\rm Mpc}}\Big)^{3} \Big(\frac{f_{b}}{1/75}\Big)^{-1} \frac{\dot{\rho}_{\rm GRB}}{0.5\,{\rm Gpc}^{-3}\,{\rm yr}^{-1}}\,{\rm yr}^{-1}.
\label{eq:Ngrb}
\end{equation}
Here $f_{b}$ is the GRB beaming factor (e.g., Gueta, Piran \& Waxman 2005),
and $\dot{\rho}_{\rm GRB}$ is their rate per unit volume.
Assuming GRB afterglows can be observed for 100\,yr (afterwards their
luminosity is equivalent to that of SN remnants), and given
the survey completeness, I conclude that there are $\ltorder 1.5$
GRB afterglows in my sample.

In any case, there are two differences between GRB afterglows and the FRB persistent sources:
The first is that while the FRB persistent sources are presumably
expanding only mildly relativistically
(Waxman 2017), GRB afterglows are expected to expand relativistically.
This may result in some differences in the angular size of FRB persistent sources
vs. GRB radio afterglows and the scintillation induced variability (e.g., different frequency-dependent variability).
Furthermore, there may be some differences in the radio spectrum.
I suggest that radio follow-up observations are required
in order to reveal the nature of the luminous radio sources reported in
this paper.

\section{Summary}
\label{sec:disc}

To summarize, I present a survey aimed at searching for luminous compact radio sources
in galaxies with $z<0.025$.

\begin{enumerate}

\item I find 11 sources with radio luminosity, at $1.4$\,GHz, of 
$\nu L_{\nu}>3\times10^{37}$\,erg\,s$^{-1}$ (i.e., $>10\%$ of the FRB\,121102 persistent source luminosity)
which are spatially associated with disks or star-forming regions of galaxies.
Here I exclude sources that are spatially coincidence with galactic centers.

\item Given the completeness estimate for the galaxy catalog, and the FIRST survey area,
I place an upper limit on the density of luminous persistent sources in
the nearby Universe ($\ltorder 5\times10^{-5}$\,Mpc$^{-3}$).
This upper limit assumes that luminous persistent sources follow the $g$-band
luminosity of galaxies.
If FRBs are related to galactic nuclei this limit will be changed to $\ltorder3\times10^{-4}$\,Mpc$^{-3}$.

\item Such luminous radio sources are rare -- about $\ltorder 10^{-3}$ per $L_{*}$ galaxy.

\item Assuming a persistent source life time of $t_{\rm age}=100$\,yr,
their birth rate
is $\ltorder5\times10^{-7}(t_{\rm age}/100{\rm yr})^{-1}$\,yr$^{-1}$\,Mpc$^{-3}$.

\item Assuming all FRBs repeat and are associated with persistent radio
sources, I set a lower limit on the FRB rate per persistent source
of $\gtorder0.8$\,yr$^{-1}$.

\item About $3\%$ of the galaxy-population integrated luminosity is in
galaxies fainter than the absolute mag. of the FRB\,121102 host ($g\approx-16.6$).
This suggests that it is too early to conclude that FRBs prefer dwarf galaxies.

\item If some of the candidates in Table~\ref{tab:Cand} are
associated with FRBs then a few-days observation with
sensitive (i.e., Parkes-like)
radio telescopes may reveal FRB events from these sources.
The detection of FRBs from such nearby galaxies may allow
us to resolve the persistent source, and to probe
the FRB luminosity function at lower luminosities
than in the case of FRB\,121102.

\end{enumerate}

\acknowledgments

I would like to thank Eli Waxman, Doron Kushnir, and Boaz Katz
for the many discussions that led to this paper,
and Orly Gnat, Barak Zackay, Eli Waxman, and Laura Spitler
for comments on the manuscript.
I would also like to thank an anonymous referee for constructive comments.
E.O.O. is grateful for support by
grants from the 
Israel Science Foundation, Minerva, Israeli ministry of Science,
the US-Israel Binational Science Foundation
and the I-CORE Program of the Planning
and Budgeting Committee and The Israel Science Foundation.

\end{document}